\documentclass[%
 reprint,
 amsmath,amssymb,
 aps,
]{revtex4-1}

\usepackage{graphicx}
\usepackage{dcolumn}
\usepackage{bm}

\begin{document}

\preprint{APS/123-QED}

\title{Gravitational solitons on Kasner background revisited: \\The simplest solitons with physical context}

\author{Christos Karathanasis}
 \email{chriskarath28@gmail.com}
\affiliation{%
 Physics Department, National and Kapodistrian University of Athens.\\
}%

\author{Theocharis Apostolatos}
 \homepage{http://users.uoa.gr/~thapostol}
  \email{thapostol@phys.uoa.gr }
\affiliation{%
 Physics Department, National and Kapodistrian University of Athens.\\
}%

\date{\today}

\begin{abstract}
We revise the one-pair complex poles soliton solutions on a Kasner background. In the literature, these were rejected as solutions with no cosmological interest due to singularities that supposedly show up at space-like infinity. The only accepted solutions of this kind were those with background metric parameter $d=\pm 1$. By computing the scalars $I, J$ we find that there are no scalar singularities at all, for a wide range of the background parameter $d$. This means that there are actually an enormous number of acceptable simple complex-poles solutions, besides the $d=\pm1$ cases. These solutions are interesting, because they are much simpler than the two-pairs complex poles solutions and, consequently, it is easier to draw conclusions and relate physical phenomena to them.
\end{abstract}

\pacs{Valid PACS appear here}
\maketitle


\section{The One-Pair Complex-Poles Soliton Solutions}

Soliton solutions in general relativity are some remarkable solutions of the Einstein equations in vacuum. The mathematical tool needed to produce such solutions was introduced by Zakharov and Belinski \cite{belinski1978}. In their pioneer paper, they demonstrated how we could use the Inverse Scattering Method (ISM) in order to integrate the nonlinear partial differential equations that result from the Einstein equations in vacuum.

The method of producing soliton solutions requires the knowledge of a suitably symmetric metric that is a vacuum solution of the Einstein field equations. This is called the background metric and the method allows one to construct new metrics that feature disturbances, propagating along an axis of symmetry. The produced metrics bear no resemblances with the background metric near the soliton origin and on the light cones. This indicates that the solitons significantly affect the background metric.

In order to be able to use this method we must choose a background metric with the desired form:
\begin{equation}\label{1}
ds^{2}_{0}=f_{0}(z,t)(dz^{2}-dt^{2})+(g_{0})_{ab}(z,t)dx^{a}dx^{b}, \ a,b=1,2.
\end{equation}

The produced solutions have the same form as the background metric. We are going to use the Kasner metric as a background. The one-pair complex-poles soliton solutions are the simplest possible solutions with physical meaning. They are given by the equations:
\begin{equation}\label{69}
(g_{1})_{11}=\sigma^{(sn)}_{1}t^{1+d}, \ (g_{2})_{22}=\dfrac{t^{2}}{(g_{1})_{11}},
\end{equation}
\begin{equation}\label{70}
f_{1}=C_{1}f_{0}\dfrac{(\sigma^{(sn)}_{1})^{1+d}(1-\cos^{2}\gamma_{1})}{(\sigma^{(sn)}_{1}+1)^2-4\sigma^{(sn)}_{1}\cos{^2}\gamma_{1}},
\end{equation}
where $\sigma^{(sn)}_{1}$ and $\gamma_{1}$ are given by the equations (4.75) and (4.76) in \cite{belinski2001gravitational}. In this paper we will examine this metric for various values of the background parameter $d$.

All these solutions are singular at $t=0$ but this singularity is of cosmological origin, since it is also present in the background metric itself. To further explore these metrics, we are going to compute the scalars $\Psi_{0,2,4}, \ I, \ J$ \cite{stephani}. To simplify the expressions of those, we are going to use the coordinate transformation which was introduced by Verdaguer \cite{verdaguer}. Using those coordinates we find:
\begin{equation}
\sigma_{1}^{(sn)}=[\tanh(\alpha T)]^{-2sn},
\end{equation}
where the label $(sn)$ stands for $(\pm)$ in \cite{belinski2001gravitational} and the constant $sn$ is respectively either equal to $+1$ or $-1$. 
The metric components (\ref{69}, \ref{70}) in the coordinates $(T,Z,X,Y)$ become:
\begin{equation}\label{74}
(g_{1})_{11}=[\cosh(2 \alpha Z) \sinh(2 \alpha T)]^{(1 + d)} [\tanh(\alpha T)]^{-2sn},
\end{equation}
\begin{equation}\label{75}
(g_{1})_{22}=[\cosh(2 \alpha Z) \sinh(2 \alpha T)]^{(1 - d)} [\tanh(\alpha T)]^{2sn},
\end{equation}
\begin{equation}\label{76}
\begin{split}
f_{1}=[\cosh(2 \alpha Z)]^{(d^2 - 1)/2} [\sinh(2 \alpha T)]&^{(d^2 +3)/
  2} \\ &\times [\tanh(\alpha T)]^{- 2sn\cdot d}.
  \end{split}
\end{equation}

In these coordinates $(T,Z,X,Y)$ the expressions for the $\Psi$-scalars are simplified \citep{belinski2001gravitational,verdaguer}:
\begin{equation}\label{77}
\begin{split}
\Psi_{2}=\dfrac{\alpha^{2}}{f_{1}}\Bigr[\dfrac{d^2 - 1}{2[\cosh(2 \alpha Z)]^{2}}+&\dfrac{ (d^2 + h^{2}-1)/2}{[\sinh(2 \alpha T)]^{2}}\\
&-h \cdot sn\dfrac{  d\cdot \cosh[2 \alpha T]}{[\sinh(2 \alpha T)]^{2}}  \Bigr],
\end{split}
\end{equation}
\begin{equation}\label{78}
\Psi_{0}= -2\dfrac{\alpha^{2}}{f_{1}}\Omega^{+}, \ \Psi_{4}= -2\dfrac{\alpha^{2}}{f_{1}}\Omega^{-},
\end{equation}
where 
\begin{equation}\label{79}
\begin{split}
\Omega^{\pm}=&-\dfrac{d (d^2 - 1)}{2}\left(1 - \dfrac{1}{2\cosh^{2}(2 \alpha Z)}\right)\\
\\& -sn \ h\left(\dfrac{h^{2}}{4}-1\right)\dfrac{\cosh^{2}[2\alpha T]}{\cosh^{2}[2\alpha Z]+\sinh^{2}[2\alpha T]} \\
&+\Bigl(\dfrac{d}{4}(1-3h^{2}-d^{2})\\
&-sn\dfrac{h}{4}(1-h^{2}-3d^{2})\cosh[2\alpha T] \Bigr) \sinh^{-2}[2\alpha T]\\
&\pm\dfrac{\sinh[2\alpha Z]}{\sinh[2\alpha T]}\Bigl[-sn\dfrac{h(h^{2}/4-1)\cosh[2\alpha Z]}{\cosh^{2}[2\alpha Z]+\sinh^{2}[2\alpha T]}\\
&+(1-d^{2})\left(-sn\dfrac{3h}{4}+\dfrac{d}{2}\right)\dfrac{\cosh[2\alpha T]}{\cosh[2\alpha Z]}\Bigr],
\end{split}
\end{equation}
where $n$ is the number of solitons, $h=\sum_{i=1}^{n}h_{i}$ and $h_{i}$ are arbitrary constants used in \cite{verdaguer}. They were introduced in order to study cases of degenerate poles. In this discussion we will mainly consider the simple case of $h_{1}=h_{2}=1$, known as the true soliton case. Note that the expression for the $\Omega^{\pm}$ given in \cite{belinski2001gravitational, verdaguer} includes an erroneous term. Namely, in the fourth row, the second term in the first parenthesis is $h^{2}$ rather than $h$, which is written in the corresponding expression in the literature.

Coordinates $(T,Z,X,Y)$ make the analytic expressions simpler. If the metric is singular at space-like ($T<Z, \ Z\rightarrow\infty$), time-like ($Z<T, \ T\rightarrow\infty$) or light-like infinities ($T=Z, \ Z\rightarrow\infty$), then the scalars $I, \ J$ will diverge at those limits. The cases of $d=\pm1$ are special, because, with a suitable coordinate transformation, the background can be shown to be the Minkowski spacetime.  For this reason we are going to examine those cases separately.

\section{Background with $d\neq\pm1$}
From equations (\ref{76}-\ref{79}), it is easy to see that the scalars $I, \ J$ do not diverge for suitable values of the background parameter $d$. At the aforementioned infinities, we see that the functions $\Omega^{\pm}$ obtain finite values. Therefore, for the behavior of the scalars, it suffices to consider only the behavior of $f_{1}$. From equations (\ref{77}, \ref{78}) we see that $\Psi_{0,2,4}\propto f_{1}^{-1}$, so when $f_{1}$ diverges, $\Psi$'s tend to 0. On the other hand, $\Psi$'s diverge when $f_{1}$ tends to 0, which corresponds to the cosmological singularity that is also present in the background. Furthermore, from equation (\ref{76}) we see that $f_{1}$ diverges when $d^{2}>1$ at space-like infinity, whereas at time-like and light-like infinities $f_{1}$ diverges regardless of the value of $d$. Consequently, for $d^{2}>1$ the scalars tend to zero at the corresponding infinities and the metric features no singularities.

The results described above have been obtained with a straightforward analysis of equations (\ref{76}-\ref{79}). However, there is one more constraint, that the background parameter needs to satisfy, so that the metric is non-singular. This extra constraint cannot be seen directly from the above equations. The constraint comes from the region of spacetime in the initial coordinates $(t,z,x,y)$where $t<<z, \ z\rightarrow\infty$, i.e. close to the $z-$axis. To discover this constraint we will need the analytic expressions for the behavior of the scalars $I, \ J$ at the aforementioned limit. $I ,\ J$ have the same behavior, written as power series in this region, so it suffices to present the results for only one of those. At space-like infinity, where $t<<z, \ z\rightarrow\infty$, we have:
\begin{equation}\label{80}
I= C t^{-d^{2}+4sn \cdot d-7}z^{-4sn \cdot d+8}+\mathcal{O}(z^{-4sn \cdot d+7}),
\end{equation}
where $C$ is some constant. The scalar will not diverge if $-4sn\cdot d+8\leq0$, that is when $sn\cdot d\geq2$ .This means that there is an extra constraint that $d$ needs to satisfy. Combining the two constraints we conclude that there is a wide class of one-pair complex-pole soliton solutions that feature no infinities all, with respect to scalar singularities. If $sn=+1$, then the accepted values of the background parameter are $d\geq2$, otherwise if $sn=-1$, then $d\leq-2$. The accepted values of $d$ depend on the sign of the pole trajectories $sn$ that we have chosen.  

It is interesting to compare at these limits the soliton solutions with the background one, that is the Kasner metric. In order to do this, we will begin with the metric itself. At space-like infinity ($t<<z, \ z\rightarrow \infty$) we have:
\begin{equation}\label{88}
\dfrac{(g_{1})_{11}}{(g_{K})_{11}}=\left(\dfrac{4z^{2}}{t^{2}}\right)^{sn}+\mathcal{O}(z^{2sn-1}),
\end{equation}
\begin{equation}\label{89}
\dfrac{(g_{1})_{22}}{(g_{K})_{22}}=\left(\dfrac{4z^{2}}{t^{2}}\right)^{-sn}+\mathcal{O}(z^{-2sn-1}),
\end{equation}
\begin{equation}\label{90}
\dfrac{f_{1}}{f_{K}}=4^{-2+sn\cdot d}t^{2(1-sn\cdot d)}z^{2sn\cdot d -4}+\mathcal{O}(z^{2sn\cdot d -5}).
\end{equation}
At time-like infinity ($z<< t, \ t\rightarrow \infty$) we have:
\begin{equation}\label{91}
\dfrac{(g_{1})_{11}}{(g_{K})_{11}}=1+2c_{1}\dfrac{sn}{t}+c_{1}sn\dfrac{c_{1}^{2}+z^{2}}{t^{3}}+\mathcal{O}(t^{-4}),
\end{equation}
\begin{equation}\label{92}
\dfrac{(g_{1})_{22}}{(g_{K})_{22}}=1-2c_{1}\dfrac{sn}{t}+4\dfrac{c_{1}^{2}}{t^{2}}-c_{1}sn\dfrac{9c_{1}^{2}+z^{2}}{t^{3}}+\mathcal{O}(t^{-4}),
\end{equation}
\begin{equation}\label{93}
\dfrac{f_{1}}{f_{K}}=\dfrac{1}{16c_{1}^{2}}+\dfrac{sn\cdot d}{8c_{1}t}-\dfrac{1}{16t^{2}}+sn\dfrac{3c_{1}^{2}+z^{2}}{16c_{1}t^{3}}+\mathcal{O}(t^{-4}).
\end{equation}
At light-like infinity ($t\approx z, \ z\rightarrow \infty$) we have:
\begin{equation}\label{116}
\begin{split}
\dfrac{(g_{1})_{11}}{(g_{K})_{11}}=&1 +2 \dfrac{c_{1}-t + z}{z}  + 
 \dfrac{sn(2c_{1}-t+z)}{\sqrt{c_{1}z}}\\ &+\mathcal{O}(z^{-3/2},(z-t)^{2}),
\end{split}
\end{equation}
\begin{equation}\label{117}
\begin{split}
\dfrac{(g_{1})_{22}}{(g_{K})_{22}}=&1 +2 \dfrac{c_{1}-t + z}{z}-\dfrac{sn(2c_{1}-t+z)}{\sqrt{c_{1}z}} \\
& +\mathcal{O}(z^{-3/2},(z-t)^{2}),
 \end{split}
\end{equation}
\begin{equation}\label{118}
\begin{split}
\dfrac{f_{1}}{f_{K}}=&\dfrac{1}{32 c_{1}^{2}}+\dfrac{d \cdot sn}{16\sqrt{c_{1}^{3}z}}+\dfrac{4d^{2}-1}{64c_{1}z}\\
&-\left(1+\dfrac{d \cdot sn\sqrt{c_{1}}}{\sqrt{z}}\right)\dfrac{-t+z}{32c_{1}^{3}}+\mathcal{O}(z^{-3/2},(z-t)^{2}).
\end{split}
\end{equation}

From these equations we can speculate that before the passage of the wave and at large $z$, which is the space-like limit, the metric is not Kasner. However, after the passage of the wave and for large values of $t$ we see that the metric tends to the background. The $g_{11}, g_{22}$ components in this limit are equal to the background values and the $f$ component is a multiple of the background value. As a consequence, it is highly likely that there is a suitable coordinate transformation that brings the metric to a Kasner form.

An invariant characterization of the metric is possible through the scalars $I, \ J$. Besides the fact that those scalars indicate whether the metric is singular or not, they are also useful because they could be used as an independent criterion to check if the metric is indeed a Kasner metric in unusual coordinates. 

The scalars for Kasner background are:
\begin{equation}\label{94}
I_{K}=\dfrac{( d^{2}-1)^{2}  }{64(3 + d^{2})^{-1}} t^{-(3 + d^{2})}, \ J_{K}=\dfrac{( d^{2}-1)^{4}}{512}  t^{-3 (3 + d^2)/2}.
\end{equation}
As we see both are functions of $t$ raised in some power. Due to this we have that:
\begin{equation}\label{96}
\dfrac{I_{K}^{3}}{J_{K}^{2}} =\dfrac{(d^2+3)^3}{( d^2-1)^2},
\end{equation}
which is constant. This is an invariant result. It holds in every coordinate system. This means that if the soliton metric tends to the background metric at a limit, then the aforementioned fraction of the scalars of the soliton metric must tend to this constant value as well. In order to avoid infinities at points where $J$ becomes zero,  we also formulate the following composite invariant:
\begin{equation}\label{97}
\tau=\dfrac{\sqrt[]{2}\xi}{\sqrt[]{1+\xi^{2}}}, \text{ where } \xi=\dfrac{I^{3}/J^{2}}{I_{K}^{3}/J_{K}^{2}}.
\end{equation}

\begin{figure}[ht]
        \includegraphics[width=0.5\textwidth]{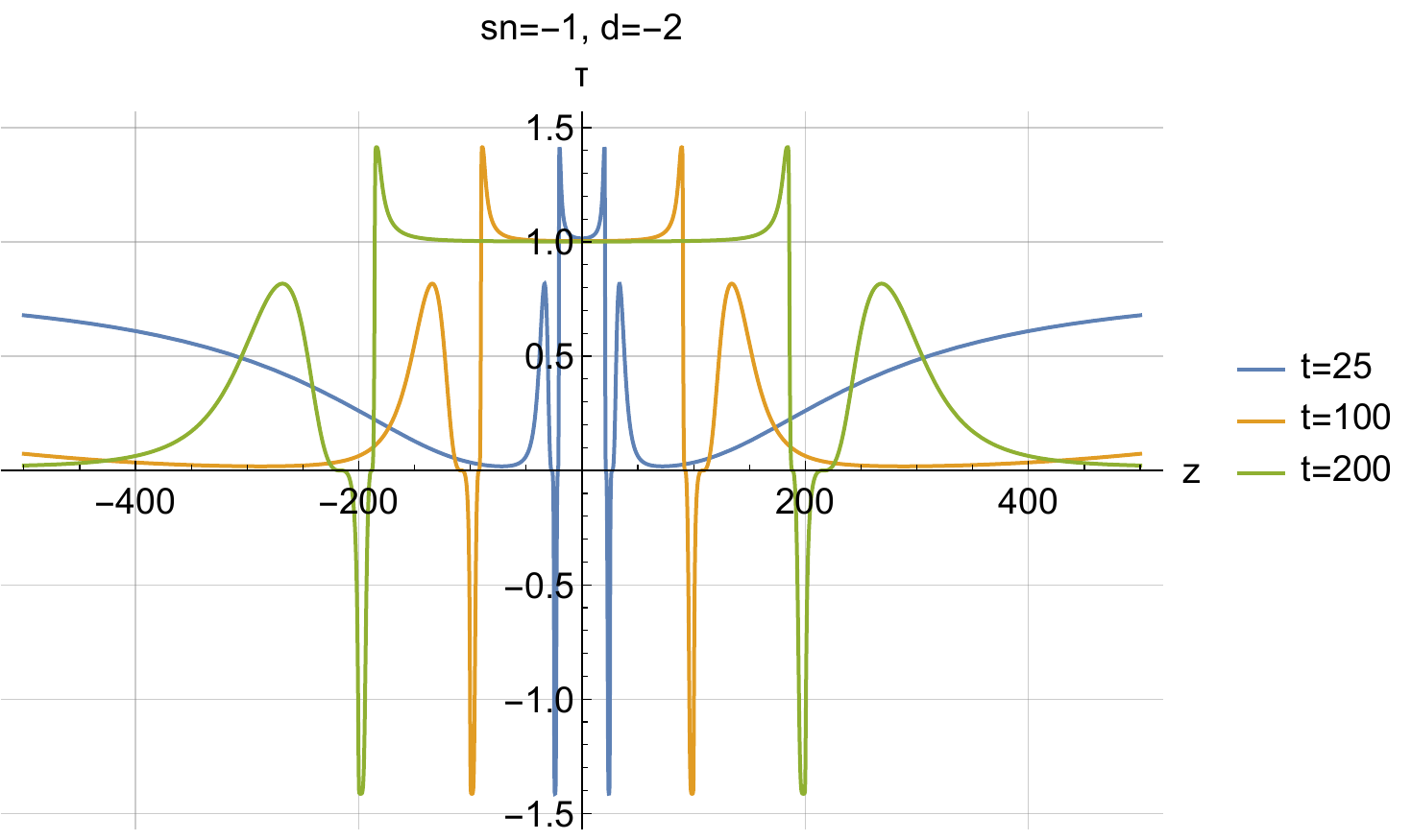}
        \caption{Plot of parameter $\tau$ at various times. Here we have chosen $sn=-1$, $c_{1}=0.1$, $z^{0}_{1}=0$ and $d=-2$. As time t passes, a growing region around $z=0$, where $\tau$ tends to unity, is that of Kasner background.}
        \label{fig4}
\end{figure}

When the soliton metric tends to the background one, then $\xi\rightarrow1$ and $\tau\rightarrow1$. When $J\rightarrow0\Rightarrow\tau\rightarrow\sqrt[]{2}$. Once again the analytic expressions are very complicated so we will only present the corresponding plots. From Fig. \ref{fig4} it can be seen that before the passage of the wave the metric is perturbated and does not tend to the background. However, after the passage of the wave $\tau\rightarrow1$, and this means that the metric indeed tends to the background in this limit.

Due to the above analysis, a suitable coordinate transformation must exist, such that the soliton metric transforms into the background in the time-like infinity. To find this we start from the metric. At time-like infinity it tends to:
\begin{equation}\label{98}
ds_{t\rightarrow\infty}^{2}=\dfrac{f_{K}}{16c_{1}^{2}}(dz^{2}-dt^{2})+(g_{K})_{11}dx^{2}+(g_{K})_{22}dy^{2}.
\end{equation}
By using the following coordinate transformation:
\begin{equation}\label{99}
\tilde{t}=c^{4/(d^{2}+3)}t, \ \tilde{z}=c^{4/(d^{2}+3)}z,
\end{equation}
\begin{equation}\label{100}
\tilde{x}=c^{-2(d+1)/(d^{2}+3)}x, \ \tilde{y}=c^{2(d-1)/(d^{2}+3)}y,
\end{equation}
where $c=(4c_{1})^{-1}$, we find that the metric can be written as:
\begin{equation}\label{101}
ds_{t\rightarrow\infty}^{2}=\tilde{t}^{(d^{2}-1)/2}(d\tilde{z}^{2}-d\tilde{t}^{2})+\tilde{t}^{1+d}d\tilde{x}^{2}+\tilde{t}^{1-d}d\tilde{y}^{2},
\end{equation}
which is indeed the Kasner metric.

From equations (\ref{91}-\ref{118}) we see that, at first order, the expressions of the metric components have the same behavior as those of the background metric at the time-like and light-like limits. However, from the above discussion, we conclude that the metric tends to the background metric only at the time-like limit. This is due to the fact that, at light-like limit, terms of higher order have more significant derivatives.  Consequently, in the limit $t \neq z\rightarrow \infty$, the derivatives of the components of the metric have different behaviors. Thus, the scalars $I, \ J$ do not tend to the background scalars.  We also see this result in Fig. \ref{fig4}, where the function $\tau$ does not tend to unity at $t=z$.

As was mentioned previously, the scalars $I, \ J$ are very useful, but they do not have a direct physical interpretation. The scalars with direct physical interpretation are $\Psi_{0}$ and $\Psi_{4}$. Due to this, it is interesting to present the plots of those. However, the figures of $\Psi_{0}$ and $\Psi_{4}$ are symmetric under the exchange $z\rightarrow-z$. Hence, we will present the plot of only one of those two.

\begin{figure}[ht]
        \includegraphics[width=0.52\textwidth]{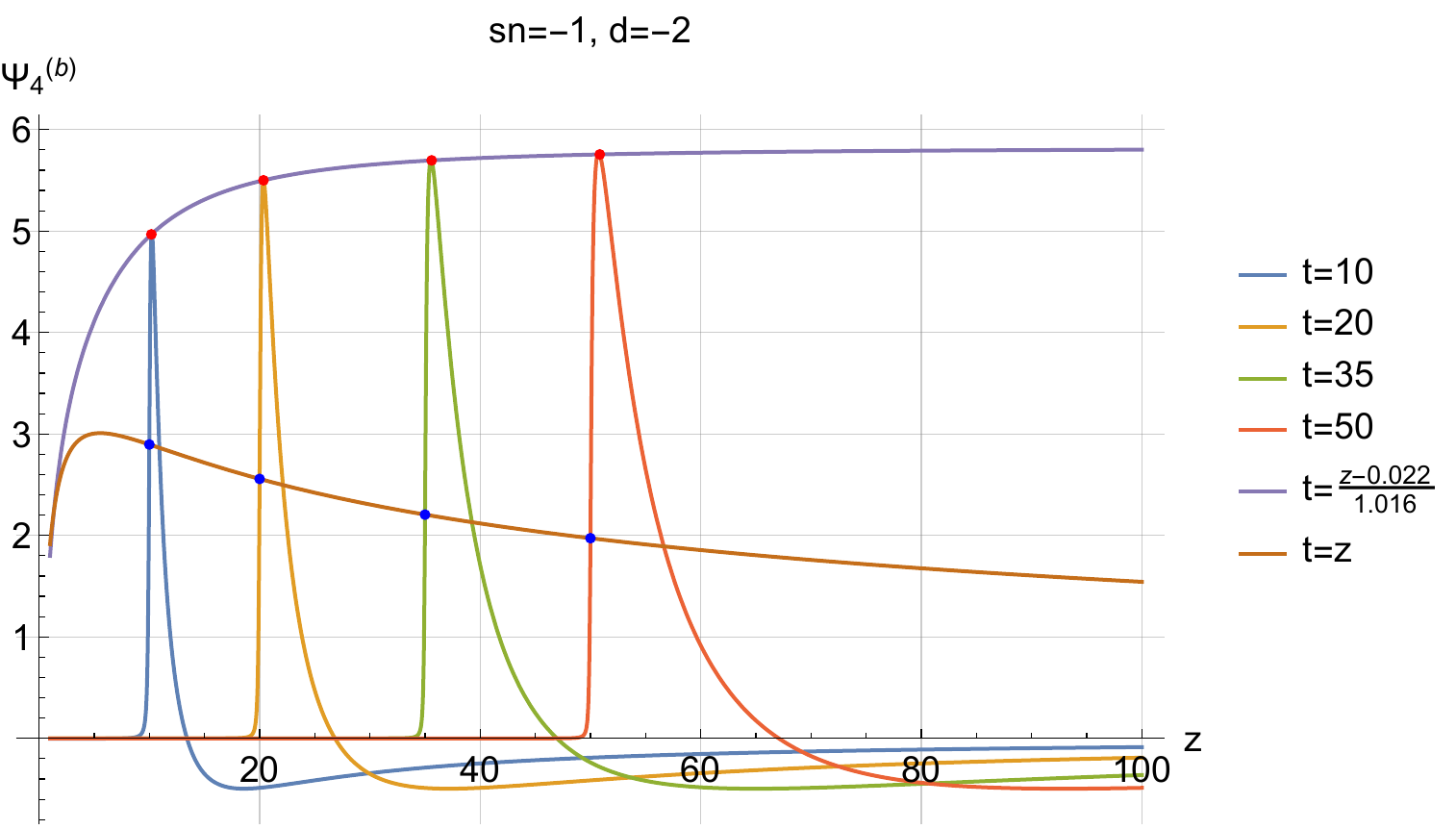}
        \caption{Plot of the scalar $\Psi_{4}^{(b)}$ with $A=t^{(d^{2}-1)/4}$ at various times. Here we have chosen $sn=-1$, $c_{1}=0.1$, $z^{0}_{1}=0$ and $d=-2$. The curve passing through the dots correspond to the motion of the peaks.}
        \label{fig19}
\end{figure}

Furthermore, we have the freedom of choosing a suitable boost $A$, when plotting the $\Psi$-scalars. In the literature, the boost that has been used is $A=\sqrt[]{f}$ when plotting $\Psi_{0}$ and $A=1/\sqrt[]{f}$ when plotting $\Psi_{4}$. With these boosts we have at light-like limit:
\begin{equation}\label{107}
\Psi_{4}^{(b)}=A^{-2}\Psi_{4}=C't^{-1/2}+\mathcal{O}(t^{-3/2}).
\end{equation}
This dependence is being referred to as typical behavior of gravitational waves \cite{belinski2001gravitational}. 

However, solitons in other aspects of physics are shape-preserving wave packets. To observe such behavior in gravitational solitons we have chosen a different $A$ to boost our tetrad. Choosing $A=t^{(d^{2}-1)/4}$ when we plot $\Psi_{0}$ or ($A=t^{-(d^{2}-1)/4}$ when we plot $\Psi_{4}$), we see from Fig. \ref{fig19} that we obtain the desired result. In this coordinates, the scalars, for small values of $t$, are increasing up to a specific time, where they acquire their maximum values. After that time, they propagate with their peaks maintaining a constant value.

It is obvious from these plot that the metric  features two disturbances.  One moves towards smaller $z$ up to $z\rightarrow-\infty$ and the other  towards greater $z$ up to $z\rightarrow+\infty$. 

From the above discussion it should be understood that the peak of the soliton is not located at spacetime points with $t=z$. The trajectory of the peaks in the $(z,t)$ plane is close to the $z=t$ line. The actual velocity of the peak slightly exceeds the speed of light (see the analytical expression for the magenta curve in Fig. \ref{fig19}).

Since the scalars $\Psi_{4}^{(b)}$ and $\Psi_{0}^{(b)}$ are symmetric, the trajectory of the peak for negative values of $z$ will be given from the same equation with the exchange $z\rightarrow-z$.

\section{Background with $d=\pm1$}

\begin{figure}
        \includegraphics[width=0.48\textwidth]{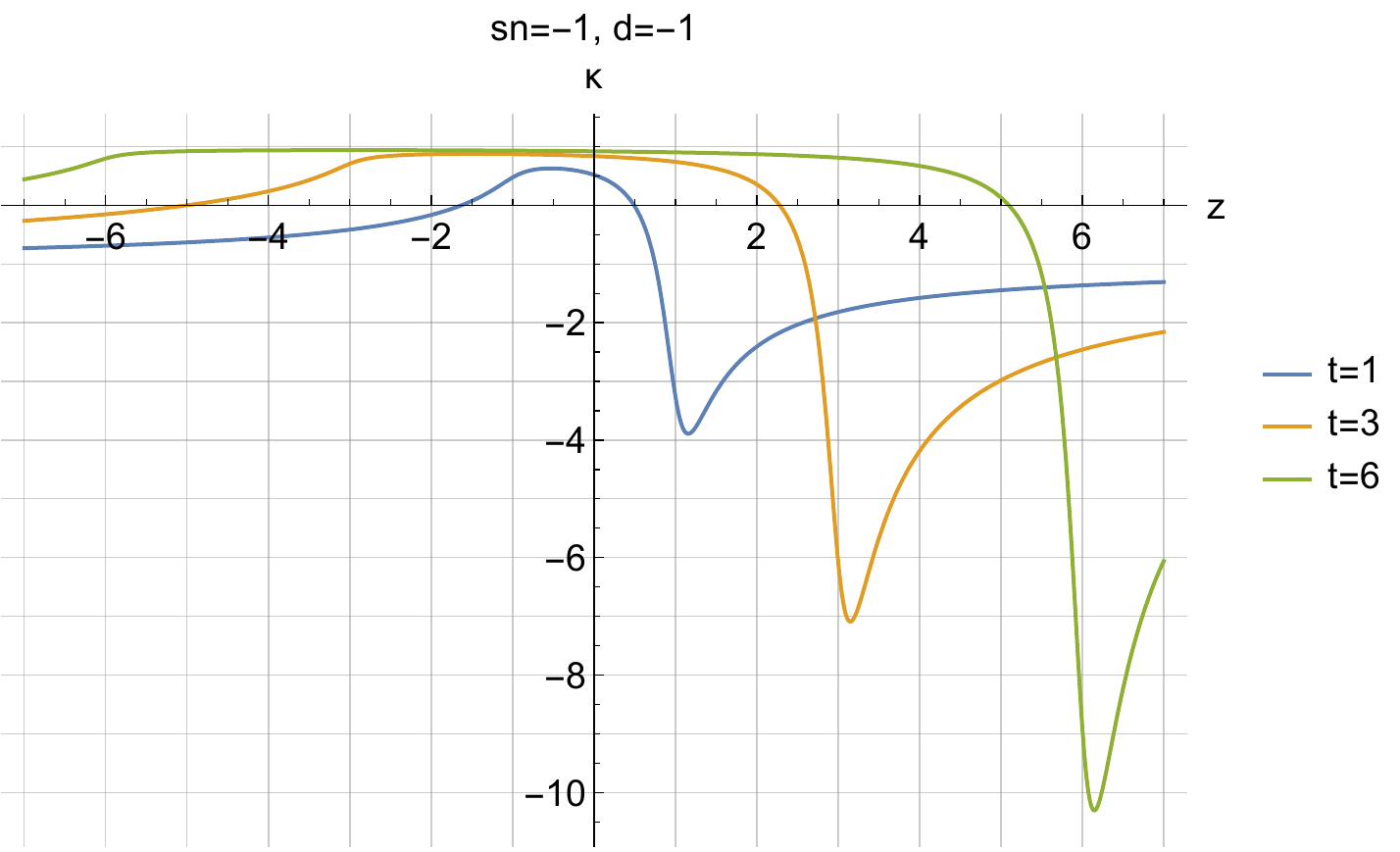}
        \caption{Plot of $\kappa$ at various times. Here we have chosen $sn=-1$, $c_{1}=0.1$, $z^{0}_{1}=0$ and $d=-1$. The function $\kappa$ tends to unity only at the time-like limit. This is the region of spacetime where the soliton metric tends to the background one.}
        \label{fig17}
\end{figure}

The cases with $d=\pm1$ are special ones because the background is the Minkowski spacetime. Again an analysis of equations (\ref{76}-\ref{79}) shows that, for $|d|=1$ there is no singularity at the infinities. However, there is a constraint that can not be seen directly from these equations. To find this constraint, we are going to use coordinates $(t, \ z, \ x, \ y)$ and expand the scalars $I, \ J$ as power series at the region of space where $t<<z, \ z\rightarrow\infty$. As mentioned before, those two scalars have the same behavior at infinities, so it is sufficient to present the results for one of those. At space-like infinity we have:
\begin{equation}\label{111}
I=12+(-18 t^2 + 64 (z^{0}_{1})^2) z^{-2}+\mathcal{O}(z^{-3}), \ \text{for} \ sn=d=-1,
\end{equation}
\begin{equation}\label{113}
I=12+(-18 t^2 + 12 (z^{0}_{1})^2) z^{-2}+ \mathcal{O}(z^{-3}), \ \text{for} \ sn=d=1,
\end{equation}
\begin{equation}\label{114}
I=Cz^{12}+\mathcal{O}(z^{11}), \ \text{for} \ sn=-d=\pm1.
\end{equation}

From these we conclude that, in order for the metric to feature no scalar singularities, we should have $d=sn=\pm1$. 

The expansions of equations (\ref{88}-\ref{118}) are also valid for $|d|=1$. Consequently, the soliton metric tends to the background only in the time-like limit. However, from equations (\ref{94}), we see that $I_{K}=J_{K}=0$ when $d=\pm1$. This means that we can not formulate the function $\tau$ of equation (\ref{97}) for these metrics. Therefore, we will use the shear and the expansion invariants alternatively \cite{stephani}.

For a metric of the form (\ref{1}), with $g_{11}=t^{2}/g_{22}$, the rotation is equal to zero while the other two scalars are given by:
\begin{equation}\label{121}
\theta=\dfrac{1}{2}\dfrac{1}{f \ t},
\end{equation}
\begin{equation}\label{122}
\sigma=\dfrac{1}{2f}\left(\dfrac{1}{t}-\dfrac{(g)_{11,t}}{(g)_{11}}-\dfrac{(g)_{11,z}}{(g)_{11}}\right).
\end{equation}

From these we find that, for the background metric, we have:
\begin{equation}
\dfrac{\sigma_{K}}{\theta_{K}}=-d.
\end{equation}

This result is again invariant. In order to see if the soliton metric tends to the background we will use the following function:
\begin{equation}\label{125}
\kappa=\dfrac{\sigma/\theta}{\sigma_{K}/\theta_{K}}=-\dfrac{\sigma/\theta}{d}.
\end{equation}

When $\kappa\rightarrow1$, we will assume that the soliton metric tends to the background. Thus, from Fig. \ref{fig17}, we conclude that the metric tends to the background at the time-like infinity. 

\begin{figure}
        \includegraphics[width=0.5\textwidth]{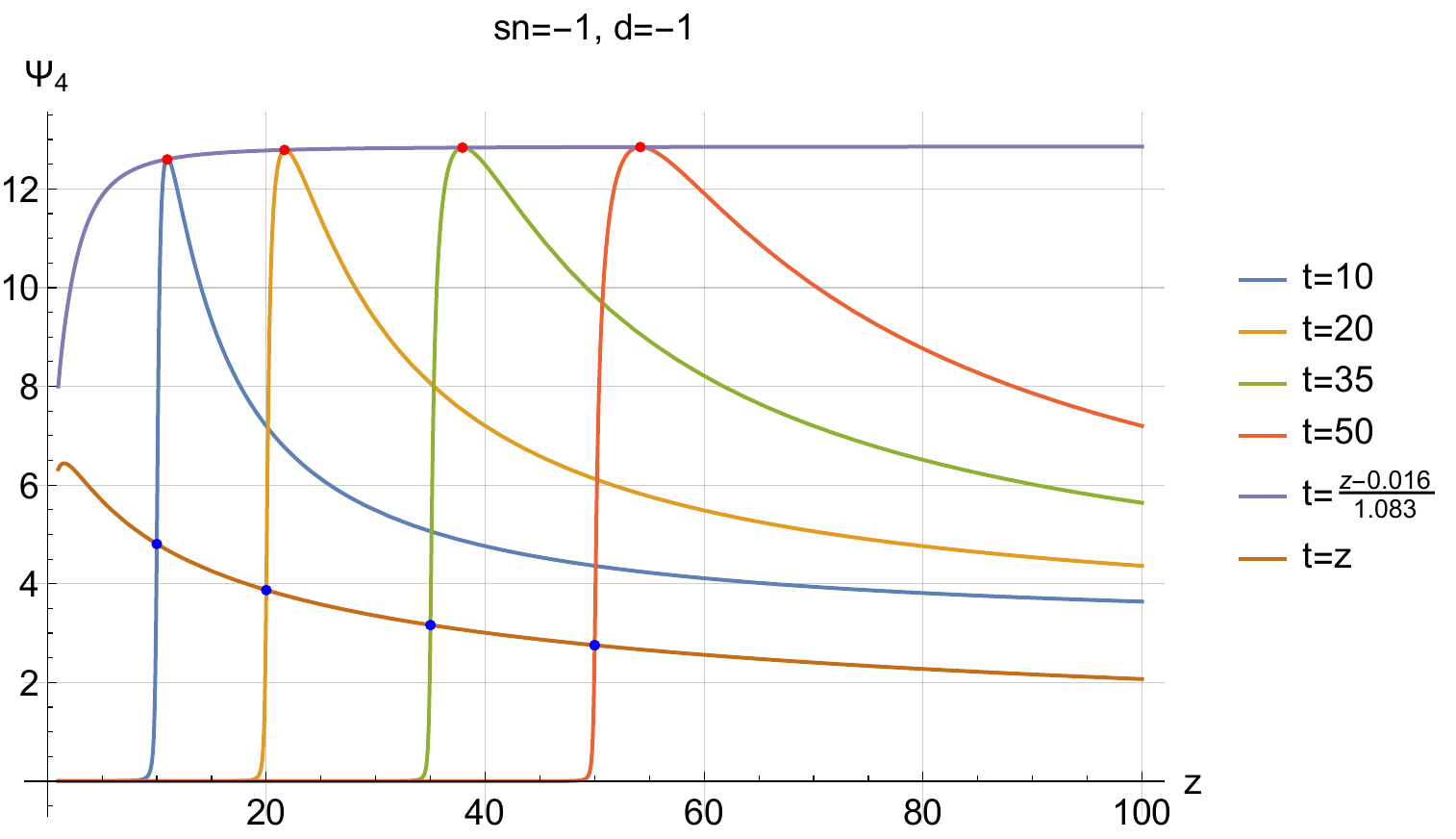}
        \caption{Plot of $\Psi_{4}$ at various times. Here we have chosen $sn=-1$, $c_{1}=0.1$, $z^{0}_{1}=0$ and $d=-1$. The curve passing through the dots correspond to the motion of the peaks.}
        \label{fig20}
\end{figure}

Finally, Fig. \ref{fig20} shows the scalar $\Psi_{4}$ for $d=sn=-1$. The boost we had used in the previous cases is equal to unity for $d=\pm 1$, therefore there is no need to use any boost at all, in order for the peaks of the solitons to obtain a finite final value. This behavior can be clearly seen at Fig. \ref{fig20}. It has the same behavior as the soliton metric with $d\neq\pm1$. The trajectory of the peak is, again, very close to the line $t=z$.

\section{Conclusions}
We have analysed the one-pair complex pole soliton solution on Kasner background, which we claim that are the simplest solutions with physical meaning. We have used a number of scalars to study the properties of these metrics.  

We found that there is a wide range of accepted soliton solutions for various values of the background parameter $d$. Specifically, we found that the accepted soliton solutions are those with either $d=sn$ or $sn \cdot d\geq2$. Note that for an arbitrary value of the parameter $h=h_{1}+h_{2}$, instead of $h=2$ on which our previous analysis is based, the above inequality for $d$, in order to obtain non-singular soliton solutions, should be recast to $sn\cdot h \cdot d\geq h^{2}$. In the literature, only the metrics with $d=\pm1$ were accepted. New scalar functions $\tau$ and $\kappa$ were introduced, in order to compare those metrics with the Kasner background. We found that the metrics tend to the background Kasner metric only at the time-like infinity, but they deviate from the background at the space-like and light-like infinities. From equations (\ref{88}-\ref{118}) it has been found that these metrics start as inhomogeneous near the soliton origins and become homogeneous at time-like and light-like infinities. However, at space-like infinity they remain inhomogeneous and this enables us to interpret these as metrics that do not describe inhomogeneities propagating on a Kasner background.

Furthermore, a boosted coordinate system, where the peaks of the solitons assume constant values as they propagate, has been found. In this boosted coordinate system, the disturbances on the graphs of $\Psi_{0,4}$ behave like solitons in other fields of physics, in the sense that they keep constant peak values as time passes.

\bibliographystyle{plain}
\bibliography{ref}
\end{document}